# Astrometry for Dynamics

A response to the call by ESA for *White Papers* for the definition of *Large* missions

**Contact details:**
Erik Høg
Skovsøen 23
2880 Bagsværd
Denmark
Tlf: +45 4449 2008
Email: erik.hoeg@get2net.dk

Erik Høg - emeritus of
Niels Bohr Institute
Juliane Maries Vej 30
2100 Copenhagen Ø
Denmark
Website: www.astro.ku.dk/~erik

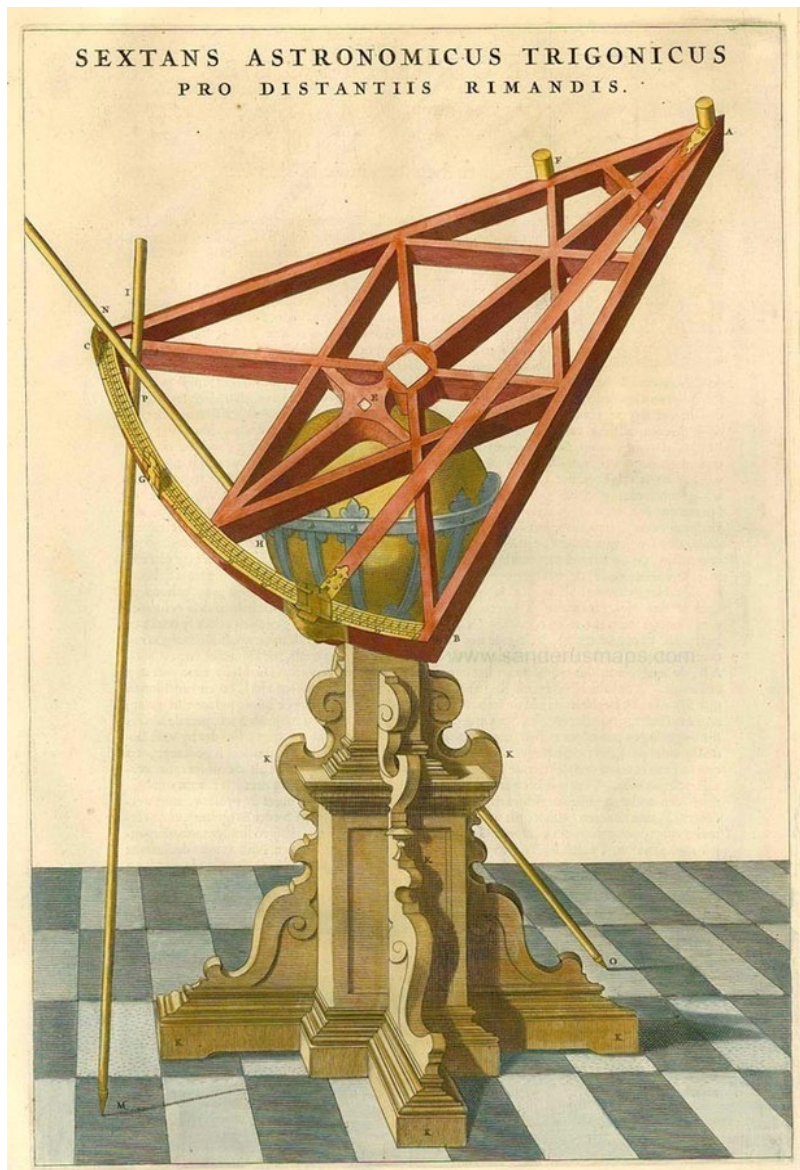

This page should list supporters for the proposal, but I will not name any because this proposal is not meant to compete with any other proposal, only to be a possible supplement.

**Acknowledgements**: I am indebted to Claus Fabricius for drawing my attention to the problem of low resolution in Gaia photometry in a correspondence about Gaia, and furthermore for mentioning the ESA call for proposals. I am also grateful to Floor van Leeuwen for his prompt answer to my question about the low resolution, quoted in the text, and to Frederic Arenou, Ulrich Bastian, Anthony Brown, Nicole Capitaine, Jørgen Christensen-Dalsgaard, Dafydd Wyn Evans, Laurent Eyer, Claus Fabricius, Gerry Gilmore, Carme Jordi, Uffe Gråe Jørgensen, Sergei Klioner, Jens Knude, Floor van Leeuwen, Dimitri Pourbaix, Annie Robin and Chris Sterken for valuable comments to a previous version. I am especially grateful to Lennart Lindegren for writing: *"I strongly encourage you to submit the proposal, which I think is definitely a very viable concept"* and *"The argument for high angular resolution photometry is compelling, for it is certainly one of the weak points of Gaia."*

**Cover figure**: Sextans Astronomicus Trigonicus by Tycho Brahe

# Astrometry for Dynamics

Proposal by Erik Høg, Niels Bohr Institute, Copenhagen, Denmark

23 May 2013

**Preample**

This proposal of a Gaia-like mission may possibly supplement the proposal expected from Anthony Brown on behalf of the GST, and it is not meant to compete. The proposal is so short because I saw the call for proposals only on 13 May while on vacation. With the deadline of 24 May for submission, a draft was distributed after a few days, and two days were left for an email discussion with colleagues who are anyway very busy, four months before the launch of Gaia.

**ABSTRACT:** Studies of the kinematic and dynamics of our Galaxy, of nearby galaxies and of our own and other planetary systems require very accurate positions obtained over long periods of time. Such studies would greatly benefit from a new Gaia-like mission. It would capitalize on the experience in Europe with Hipparcos and Gaia and on the results from these missions. Proper motions with much smaller formal errors than from Gaia alone can be obtained for one billion stars from the positions observed by Gaia and a new mission. Equally important, these proper motions will be much less affected systematically by motions in the very frequent unresolved binaries. Studies of our Galaxy and nearby galaxies will benefit from the proper motions. The study of orbits for over 100,000 objects in the solar system will profit greatly when it can be based on the positions from two global astrometry space missions by ESA. It is proposed that the photometry with low-dispersion spectra is replaced by filter photometry in 3 or 4 bands. This will provide photometry of all stars, sufficient for the required chromatic corrections of astrometry and it will give photometry of narrow double stars which cannot be obtained from the ground nor can it be obtained with Gaia because the long spectra of the two stars overlap. Ground-based surveys of multi-colour photometry and spectra will be available for astrophysical studies for a large fraction of the stars.

**Infrared and Nanoarcseconds missions**

Infrared and Nanoarcseconds missions for astrometry are worth considering, but I will leave the detailed discussion to others. An infrared mission concept, JASMINE, has been studied by Japanese colleagues for more than ten years and a small version, Nano-JASMINE is due for launch in November 2013 and has a 5 centimeter primary mirror. But Nano-Jasmine is NOT infrared, it is a CCD-based version of Hipparcos (scanning satellite on an Earth-centered circular orbit). It is the first of a planned series of three satellites of increasing size: Nano-JASMINE, Small-JASMINE, and JASMINE. The two last satellites shall observe in the infrared, allowing for better observation towards the center of the Galaxy.

It is my impression from the discussion I have seen or heard of such missions that they have great potential. I consider the nanoarcseconds region to be an important long-term goal for space astrometry in the effort to serve the needs of astrophysics. But it would be too risky to focus only on this goal for a mission already around 2030 in view of all the technical challenges, known as well as unknown ones. The astrometric community cannot soon focus on the development of a difficult new mission because the Gaia mission will take the attention in the years to come. Instead of such a single goal for astrometry, ESA should capitalize on the experience from Hipparcos and Gaia and plan for a Gaia-like mission about 2030. This would provide excellent and unique science data as I shall outline below and the mission cost may be lower because of the experience with Gaia. The mission study should include a consideration of some infrared detection capability in an otherwise Gaia-like mission in order to get closer to the center of the Galaxy. All this would require that ESA plans for astrometry on a time scale of 40 years.

**Gaia-like mission**

A Gaia-like mission is proposed in response to the call by ESA (2013). It should have a scientific performance as expected for Gaia, according to the section included below, but exceeding this where further considerations during the coming years will show the technical possibilities combined with deeper consideration of the scientific wishes. At present, two improvement are pointed out: (1) An observing epoch about 2031 for an L2 mission would provide large epoch differences from the Hipparcos mission about 1991 and the Gaia mission about 2016, (2) the proposed introduction of filter photometry instead of low-dispersion spectra could provide 3 or 4-colour photometry of all stars, especially also of narrow double stars.

An epoch difference of 15 years from Gaia means that the proper motions derived from the positions obtained at each epoch will have formal standard errors 7.5 times smaller than those from Gaia alone. This follows from the performance figures given on the Gaia website. The parallaxes from both missions will also improve because the proper motions will no longer interfere with the parallaxes.

Proper motions derived from positions observed in a short interval of time, e.g. from one space motion are very often affected by systematic errors due to the motions in unresolved binaries, it therefore takes time to obtain very good proper motions, e.g. two or more space missions. The systematic error from the orbital motions depends on the orbital period of the pair.

Unresolved double star can be discovered from astrometric observations in a single mission by the large residuals of the standard solution of linear motion for a single star. Many such discoveries were obtained with Hipparcos, in fact only 80 per cent of the stars could be solved as single stars without problems (see section 6.5.1 in ESA 2000). Many more will be problematic in Gaia observations which will thus lead to new discoveries.

With two missions a very large fraction of stars can be discovered as binaries from the residuals of either mission and from a comparison of the proper motions from each mission and from the motions derived from the positions at the two epochs. Furthermore, the acceleration in the orbit can be determined. The additional use of Hipparcos results will lead to further discoveries, depending on the orbital periods.

Many binaries and planetary systems will be discovered with these methods (see sections 1.5 and 1.6.2, respectively, in ESA 2000). What has changed significantly since Gaia was approved by ESA is the situation of the extrasolar planets. Rather than a few tens, there are now a few thousands already identified, essentially, by Kepler. Furthermore, the final Gaia epoch precision is somehow worse than what was originally asked by the scientific community. So, the expectation in terms of planet discovery and orbit fitting should be lowered. However, the discovery and characterisation of binary and multiple stars remain as valuable as they were ten years ago. A space based proper motion catalogue derived from the two Gaia missions, 15 years apart, for 1 billion stars would supersede any ground base catalogue ever compiled.

The benefit of the much more accurate proper motions for studies of the Galaxy deserves to be elaborated. The same is true for the observations of star positions in nearby galaxies and of solar system objects.

From the scientific point of view, the impact of a significantly higher accuracy of absolute proper motions can be seen in studies of galactic dynamics (probing much deeper into the galaxy as well as into the halo), dynamics of star clusters and OB associations (conditions of star formation, cluster formation), proper motions and orbits of globular clusters (galactic potential), apparent proper motions of QSOs, resulting from the orbital acceleration of the solar system in our galaxy and the aberration effects resulting from the velocity vector of the solar system in space. Direct comparisons with the Gaia results would also provide statistics on orbital binary systems from proper motion disturbances.

**Photometry**

There are two reasons to perform photometry of stars with the mission itself: (1) enable chromatic corrections of the astrometric observations, and (2) provide astrophysical information for all objects, including astrophysical classification (for instance object type such as star, quasar, etc.) and astrophysical characterisation (for instance interstellar reddenings and effective temperatures for stars, photometric redshifts for quasars, etc.).

For a new mission the first reason is still valid to the same extent as for Gaia, but the justification for the second purpose has changed. By the time of a new mission, multi-colour photometry will be available for all the observed stars with higher accuracy and better spectral resolution than the mission itself can provide. Such photometry will be provided and be available from large surveys as Pan-STARRS and LSST, providing five or six spectral bands from 300 to 1100 nm. The angular resolution of these surveys could not be found, but since they are ground based, the resolution is hardly better than 0.5 arcsec. Also spectra can be obtained in the millions by new spectrographs.

For Gaia the angular resolution along scan of the astrometric observation is about 0.12 arcsec (FWHM of the sampled line-spread function). For photometry however the low-dispersion spectra limit the resolution greatly for double and multiple stars because the spectra of the two stars will overlap, each spectrum having a length of 1-2 arcsec. This was a penalty of going from filter photometry to spectrophotometry proposed by the industrial contractor in 2005.

This is not apparent from the Gaia information quoted below: *"Photometric observations will be collected with the photometric instrument, at the same angular resolution as the astrometric observations and for all objects observed astrometrically"*. To my recent question about this issue Floor van Leeuwen replied: *"For the spectra there are limitations in densely populated areas, although we will try as far as we can to de-blend poluted spectra. So, in principle the data are there, but they may not be easy or even possible to reduce."* – This confirms my understanding of the problem of Gaia photometry for double and multiple stars.

For a new mission, it is proposed to return to filter photometry. The two prisms should be replaced by two filters which may, e.g., cut out 330-460 and 460-600 nm corresponding to B and V. The intensity in the band beyond 600 nm may be obtained from the intensity of the wide-band 330-1050 nm measured in the astrometric field by subtracting the intensities of the observed B and V, and the result may be called R. This is just an example and the exact choice of band need careful study, but the basic idea is that 3-band photometry would be obtained: B, V, R, but three bands are probably sufficient for the chromatic correction of astrometric observations.

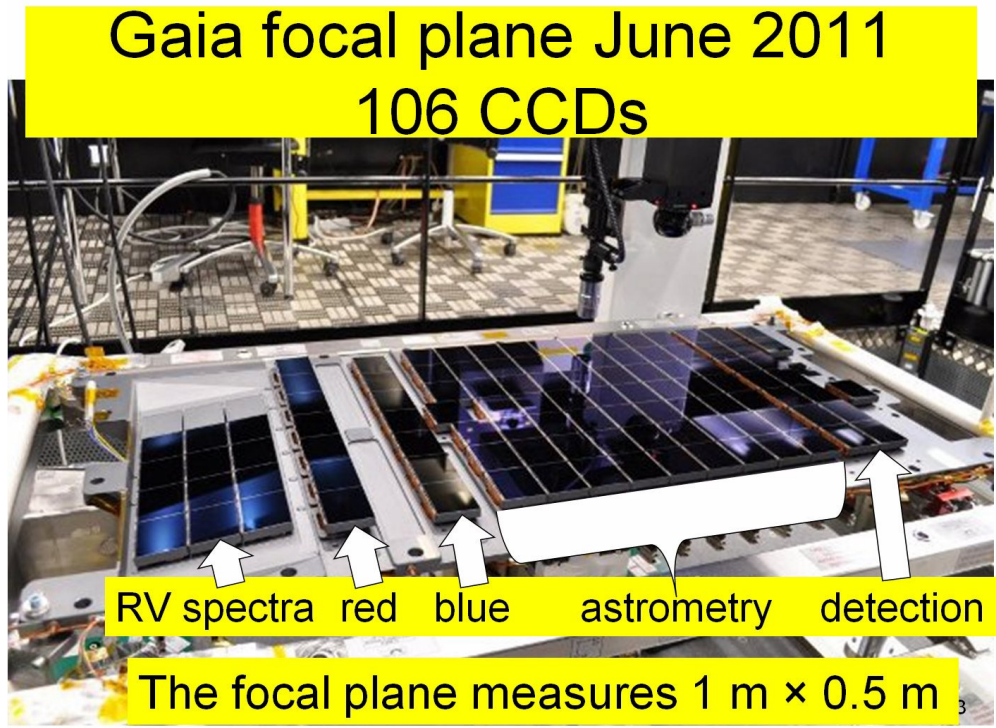

**Figure**  The focal plane of Gaia with 106 large CCDs. The stars enter from right, are detected and then measured during the passage of various CCDs. At first comes astrometry on 9 CCDs, then photometry on two CCDs measuring a short spectrum in blue, BP, and in red, RP. At left the stars are observed at high dispersion around the red Ca-triplet in order to obtain the radial velocity of the star and the intensity in the spectrum.

It should be considered to include 3 CCDs for photometry instead of two. It appears from the figure that the space would be available without any shift of the CCDs for astrometry and RV. If this is done 4-band photometry could be obtained, B, V, R, I, where one of the bands results from subtraction of the three bands from the photometry obtained in the wide astrometric band.

The accuracy of such filter photometry will be much better than the Gaia photometry from the short spectra in two respects: (1) All stars, especially the faint ones, will be less affected by noise from readout, background and parasitic stars, and (2) for components of double and multiple stars the improvement will be most pronounced, and they will obtain good photometry in many cases where Gaia could not give anything. The further advantage of the simpler data reduction is worth mentioning.

It must be emphasized that the 4-colour photometry is meant for use to make the chromatic astrometric corrections and for astrophysics when nothing else is available. Photometry was an historical necessity, when spectra were too expensive to obtain. Dedicated spectrographs are now being built which will obtain millions of stellar spectra. That provides the quantitative science. Superb multi-color all-sky IR data are already available, and the multi-colour optical surveys are being rapidly completed. It does not make sense to duplicate those with an astrometric mission.

**Origins of Gaia**

In October 1993 a proposal was submitted to ESA to study for astrometry *"a large Roemer option and an interferometric option"*, GAIA. They should be studied as two concepts for an ESA Cornerstone Mission for astrometry *"without a priori excluding either"*, as the cover letter said. Interferometry was completely dropped in January 1998, and further design was based on the Roemer concept with direct imaging on CCDs using Time Delayed Integration (TDI), a technique not previously used in space. Therefore, Gaia is a large Roemer option. – The history of the events before and after 1993 are briefly reviewed by Høg (2013).

**References**

ESA 2000, GAIA – Composition, Formation and Evolution if the Galaxy. Concept and Technology Study Report. ESA-SCI(2000)4.

ESA 2013, ESAs call for White Papers for the definition of the L2 and L3 missions in the ESA Science Programme at `http://sci.esa.int/Call-WP-L2L3`.

Gaia 2013, The website at:
`http://www.rssd.esa.int/index.php?project=GAIA&page=Science_Performance`

Høg E. 2013, Origins of Gaia. `https://dl.dropbox.com/u/49240691/GaiaOrigins.pdf`

## Science Performance of Gaia

Copied from the Gaia website in 2013

Gaia will perform micro-arcsecond (µas) global astrometry for all ~1,000 million stars down to G ≈ 20 mag — except for the ~6,000 brightest stars in the sky — by linking objects with both small and large angular separations in a network in which each object is connected to a large number of other objects in every direction. Over the five-year mission lifetime, a star transits the astrometric instrument on average ~70 times, leading to ~630 CCD transits. Gaia will not exclusively observe stars: *all* objects brighter than G ≈ 20 mag will be observed, including solar-system objects such as asteroids and Kuiper-belt objects, quasars, supernovae, multiple stars, etc. The Gaia CCD detectors feature a pixel size of 10 µm (59 milli-arcsecond) and the astrometric instrument has been designed to cope with object densities up to 750,000 stars per square degree. In denser areas, only the brightest stars are observed and the completeness limit will be brighter than 20$^{th}$ magnitude.

Photometric observations will be collected with the photometric instrument, at the same angular resolution as the astrometric observations and for all objects observed astrometrically, to:

- enable chromatic corrections of the astrometric observations, and
- provide astrophysical information for all objects, including astrophysical classification (for instance object type such as star, quasar, etc.) and astrophyscial characterisation (for instance interstellar reddenings and effective temperatures for stars, photometric redshifts for quasars, etc.).

Spectroscopic observations will be collected with the spectroscopic instrument for all objects down to $G_{RVS} \approx 16$ mag, to:

- provide radial velocities through Doppler-shift measurements using cross-correlation (~150 million stars);
- provide astrophysical information, such as interstellar reddening, atmospheric parameters, and rotational velocities, for stars brighter than $G_{RVS} \approx 12$ mag (~5 million stars); and
- provide element abundances for stars brighter than $G_{RVS} \approx 11$ mag (~2 million stars).

The spectroscopic instrument has been designed to cope with object densities up to 36,000 stars per square degree. In denser areas, only the brightest stars are observed and the completeness limit will be brighter than 16th magnitude.

In the scientific performance assessments for Gaia, all known instrumental effects are included under the appropriate in-flight operating conditions (temperature, CCD operating mode, etc.). All error sources are included as random variables with typical deviations (as opposed to best-case or worst-case deviations). All performance estimates include a 20% contingency margin. This margin is a DPAC science margin, neither meant for nor available to the Gaia industrial prime contractor. The science margin is assumed to cover, among others:

- "scientific uncertainties" in the on-ground data analysis, including uncertainties related to relativistic corrections, aberration corrections, and the spacecraft and solar system ephemeris;
- scientific effects such as the contribution to the astrometric error budget from the mismatch between the actual and the calibrating point spread function, estimation errors in the sky background and total detection noise values that need to be fed to the centroiding algorithm, etc.;
- the fact that the sky does not contain, as assumed for the performance assessments, "perfect stars" but "normal stars", which can be photometrically variable, have spectral peculiarities such as emission lines, have unrecognised companions, be located in crowded fields, etc.;
- other astronomical environmental factors such as, e.g., localised enhanced sky-background surface brightness, unrecognised small-scale sky-background-brightness gradients, unrecognised prompt particle events, etc.

Here follow on the Gaia website (Gaia 2013) the detailed performances for astrometry, photometry and spectroscopy.

+++++++++++++++++++++++++++++++++++++++++++++++++++++